\documentclass{emulateapj}
\usepackage{graphicx}
\newcommand{\bl}[1]{\mbox{\boldmath$ #1 $}}

\slugcomment{To appear in ApJ Letters}
\shorttitle{Episodic Accretion Bursts}
\shortauthors{E. I. Vorobyov and Shantanu Basu}
\begin{document}

\title{The Origin of Episodic Accretion Bursts in the Early Stages
of Star Formation}

\author{E. I. Vorobyov\altaffilmark{1,}\altaffilmark{2},
Shantanu Basu\altaffilmark{1}}
\altaffiltext{1}{Department of Physics and Astronomy, University of Western Ontario,
London, Ontario,  N6A 3K7, Canada; vorobyov@astro.uwo.ca, basu@astro.uwo.ca }
\altaffiltext{2}{Institute of Physics, Stachki 194, Rostov-on-Don, Russia}

\begin{abstract}
We study numerically the evolution of rotating 
cloud cores, from the collapse of a magnetically supercritical core 
to the formation of a protostar and the development of a 
protostellar disk during the main accretion phase.
We find that the disk quickly becomes unstable to the development of
a spiral structure similar to that observed recently in AB Aurigae. 
A continuous infall of matter 
from the protostellar envelope makes the protostellar disk unstable, leading to spiral arms
and the formation of dense protostellar/protoplanetary clumps within them.
The growing strength of spiral arms and ensuing redistribution of mass
and angular momentum creates a strong centrifugal disbalance in the 
disk and triggers bursts of mass accretion during which the dense 
protostellar/protoplanetary clumps fall onto the central protostar.
These episodes of clump infall may manifest themselves as episodes of vigorous 
accretion rate ($ \ge~10^{-4}~M_\odot$~yr$^{-1}$) as is observed in 
FU Orionis variables. 
Between these accretion bursts, the protostar is characterized by a low 
accretion rate ($ < 10^{-6} M_\odot$~yr$^{-1}$). During the phase 
of episodic accretion, the mass of the protostellar disk remains  
less than or comparable to the mass of the protostar.
\end{abstract}


\section{Introduction}
In our present view of low-mass star formation, a protostar in the early stages
of mass assembly (in the so-called class~0 and class~I phases) is surrounded by a 
protostellar disk which is in turn deeply embedded in an infalling envelope
left over from the collapse of a rotating prestellar cloud core. 
The observed low luminosity of these protostars implies a low mass accretion rate,
and hence a long lifetime in order to achieve typical final stellar masses;
however, this is inconsistent with the
number of known class~0 and class~I objects \citep{Kenyon2}.
A possible explanation  is that the protostellar accretion proceeds in two
co-existing phases \citep{Kenyon}. Accretion from the envelope onto the protostellar disk takes
place in a fairly uniform (though generally declining in time) manner, whereas
accretion from the protostellar disk onto the central protostar occurs 
primarily in short (and infrequently observed) but
powerful episodes during which $0.01-0.1~M_{\odot}$ can be accreted. These
episodes of vigorous accretion $\dot{M} \ge 10^{-4}~M_{\odot}$~yr$^{-1}$
manifest themselves as FU Orionis variables (FU Ori). Between these accretion 
bursts, a typical class~0/class~I protostar is characterized by a low 
accretion rate $\dot{M} \sim 10^{-7}~M_\odot$~yr$^{-1}$.

The nature of FU Ori disk accretion bursts has been widely debated. 
For instance, close encounters in binary systems may cause a strong perturbation
in protostellar disks and drive high accretion rates during a relatively
short period \citep{Bonnell}. This mechanism requires rather eccentric
orbits of binary systems and obviously fails to explain FU Ori outbursts in 
isolated protostars.
An alternative idea is that the thermal instability (namely, the steep dependence of 
the disk opacity on temperature between $\sim 3000$~K and $\sim 10^4$~K)
of the optically thick innermost regions of circumstellar disks 
triggers FU Ori eruptions \citep{Lin,Clarke,Bell}. Most thermal instability models
exploit the $\alpha$-prescription of \citet{Shakura}, who suggested that
the disk effective viscosity is proportional to its temperature; an increase in disk temperature
causes a higher rate of mass accretion due to an elevated viscous mass transfer and vice versa.
Unfortunately, many aspects of this mechanism of FU Ori outbursts are completely
dependent upon the unknown value of disk effective viscosity, which determines the timescale
for outbursts.

It is known from theoretical and numerical studies that protostellar disks may be 
subject to the development
of global spiral instabilities. 
For instance, \citet{LB} have studied the nonaxisymmetric evolution
of protostellar disks and found that they are susceptible to a series of spiral
instabilities, with the fastest growing modes being the one-armed ($m=1$) and 
two-armed ($m=2$) patterns. 
Recent numerical simulations of star cluster formation \citep{Bate2} confirmed that the protostellar disks formed
around protostars were gravitationally unstable and prone to the development
of spiral density waves. 
Simulations of marginally unstable protoplanetary disks 
show the formation of flocculent and clumpy spiral structure, 
suggesting a possible transient rise in the mass accretion rate 
associated with clump infall \citep{Boss2}.
\citet{Mejia} have also reported three-dimensional simulations 
which demonstrate a single FU-Ori-like outburst associated with
the growth of spiral structure in an isolated protoplanetary disk.
Recent observations do reveal a flocculent spiral structure in
the protostellar disk of AB~Aurigae \citep{Fukagawa,Sargent}.

In this Letter, we present the first model of 
cloud core collapse which self-consistently
generates multiple accretion bursts that can be identified with 
FU Ori eruptions. The protostellar disk in our model is formed as a
result of the collapse and is not isolated from the parent core envelope.
The details of the numerical model are described in \S~\ref{model}. 
The results of simulations are presented in \S~\ref{res}. Our main
conclusions are summarized in \S~\ref{concl}.

\section{Model description}
\label{model}

We model the collapse of a rotating cloud core which is threaded by a 
frozen-in magnetic field with spatially uniform mass-to-flux ratio.
The magnetic field effect is comparable to but weaker than gravity, 
so that the core is magnetically supercritical. A magnetically-diluted
gravitational collapse ensues. Our initial core model is 
a good approximation to the supercritical cores that result from the 
fragmentation of magnetized clouds that are initially either 
critical/subcritical or supercritical \citep[e.g.,][]{Basu2,Basu3}. 

We follow the collapse through the prestellar collapse phase and into the
accretion phase which sees the development of a protostar and a 
protostellar disk. The thin-disk approximation is used in the form
appropriate for a supercritical core with
mass-to-flux ratio $\mu$ that is spatially uniform \citep{Basu2,Nakamura,Shu}.
The magnetic field pressure
enhances the gas pressure and the magnetic tension dilutes the effect
of gravity. 
The disk is symmetric about the midplane,
with a vertical magnetic field component $B_{\rm z}$ inside the disk and
both vertical and tangential components outside it. The material external to the disk
is current-free ($j=0$). We characterize the field strength 
by the parameter $\alpha \equiv \mu^{-1} = B_{\rm z}/(2 \pi {\sqrt G} \Sigma)$, 
where $\Sigma$ is the gas surface density; note $\alpha < 1$ for a 
supercritical core.

The magnetohydrodynamic equations we use are 
written in the thin-disk approximation as 
\begin{equation}
{\partial \Sigma \over \partial t} = - {\bl \nabla} \cdot (\Sigma \,{\bl v}),
\label{contin}
\end{equation}
\begin{equation}
\Sigma {d{\bl v} \over dt}= -{\bl \nabla} {\bar P} - {Z\over 4 \pi} {\bl\nabla}
B_{\rm z}^2 - (1-\alpha^2)\,\Sigma\, {\bl \nabla} \Phi,
\label{motion}
\end{equation}
where ${\bl v}$ is the gas velocity in the disk plane, $\bar P$ is the
vertically integrated gas pressure, 
and $\Phi$ is the gravitational potential.
Equation (\ref{motion}) contains the  
Lagrangian derivative $d/dt=\partial /\partial t + {\bl v}\cdot \nabla$.

We assume a two-component equation of state  ${\bar P}=c_{\rm s}^2\Sigma
+ c_{\rm s}^2 \Sigma_{\rm cr} (\Sigma/\Sigma_{\rm cr})^\gamma$, where $c_{\rm
s}$ is the isothermal speed of sound
and  $\Sigma_{\rm cr}$ is the critical gas surface density above which
the disk becomes optically thick. This
equation of state allows for a smooth transition between the isothermal
and non-isothermal regimes during the collapse. 
We use a canonical value of the critical gas volume density $n_{\rm cr}=10^{11}$~cm$^{-3}$ \citep{Larson}, 
which is equivalent to $\Sigma_{\rm cr}=36.2$~g~cm$^{-2}$  for
the gas disk in vertical hydrostatic equilibrium.
In equation~(\ref{motion}), we assume a cloud scale height $Z=c^2_{\rm s}/(\pi
G \Sigma)$ for $\Sigma\le \Sigma_{\rm cr}$ and $Z=c^2_{\rm s}/(\pi G\Sigma_{\rm
cr})$ for $\Sigma > \Sigma_{\rm cr}$.
We adopt the ratio of specific heats $\gamma=7/5$ for the optically
thick regime, appropriate for an adiabatic diatomic gas.

Equations~(\ref{contin}) and (\ref{motion}) are numerically solved in polar
coordinates ($r,\phi$) using the method of finite differences with a time-explicit,
operator-split solution procedure 
similar to that described by Stone and Norman in their ZEUS-2D code \citep{SN}.
The details of the code and results of the tests will be given in a follow-up
paper.
The numerical grid has $256\times 256$ points, which are logarithmically spaced 
in $r$-direction allowing for a good resolution in the inner cloud regions. 
The innermost grid point is located at 10~AU and the size of the first
adjacent cell is 0.3~AU.
The Truelove criterion \citep{Truelove} is preserved throughout the simulations 
-- the size of grid cells is smaller than the Jeans length.
We introduce a ``sink cell'' at $r<10$~AU, which represents 
the central protostar plus some circumstellar disk material, and impose a free inflow
inner boundary condition. 
We assume that the matter is cycled through the circumstellar disk and onto
the protostar rapidly enough so that the mass infall
through the sink cell is at least proportional to the mass accretion 
rate onto the protostar. 
We impose the outer boundary condition such that the gravitationally bound
cloud has a constant mass and volume. 
The gravitational potential of the cloud is evaluated numerically using the fast Fourier
transform \citep{BT}. 

\section{Results}
\label{res}
We have studied many different initial cloud configurations and present here a
prototype magnetized ($\alpha=0.45$) rotating cloud with mass $M_{\rm cl}=2.45~M_\odot$ and an
outer radius $r_{\rm out}=20000$~AU. The cloud is composed
of molecular hydrogen with a $10\%$ admixture of atomic helium and is initially
isothermal at $T=10$~K ($c_{\rm s}=0.188$~km~s$^{-1}$).
The initial surface density and angular velocity distributions are those
characteristic of a collapsing axisymmetric supercritical core \citep{Basu2}:
\begin{equation}
\Sigma={r_0 \Sigma_0 \over \sqrt{r^2+r_0^2}},
\label{dens}
\end{equation}
\begin{equation}
\Omega=2\Omega_0 \left( {r_0\over r}\right)^2 \left[\sqrt{1+\left({r\over r_0}\right)^2
} -1\right].
\end{equation}
Here, $\Sigma_0=3.55\times10^{-2}$~g~cm$^{-2}$ and $\Omega_0=0.5$~km~s$^{-1}$~pc$^{-1}$ 
are the central surface density and angular velocity, respectively. 
We choose a value  $r_0=c_{\rm s}^2/(1.5 G \Sigma_0)$, so that
$r_0$ is comparable to the Jeans length of an isothermal sheet. 
We mimic the slight nonaxisymmetry in more realistic models of core formation
\citep{Basu3}  by substituting $r^2$ in Eq.~(\ref{dens}) with
$r^2 ({\cos^2\phi/a^2}+a^2 \sin^2\phi )$, where the parameter $a=0.98$ denotes the
cloud oblateness.
The ratios of rotational, magnetic, and
thermal energies to the gravitational energy of the cloud are $0.4\%$, $30\%$, and $27.2\%$, respectively.
Thus, the initial cloud is gravitationally unstable. We emphasize that
our qualitative results are insensitive to the particular choice of initial conditions.

\begin{figure*}
 \centering
  \includegraphics[width=15 cm]{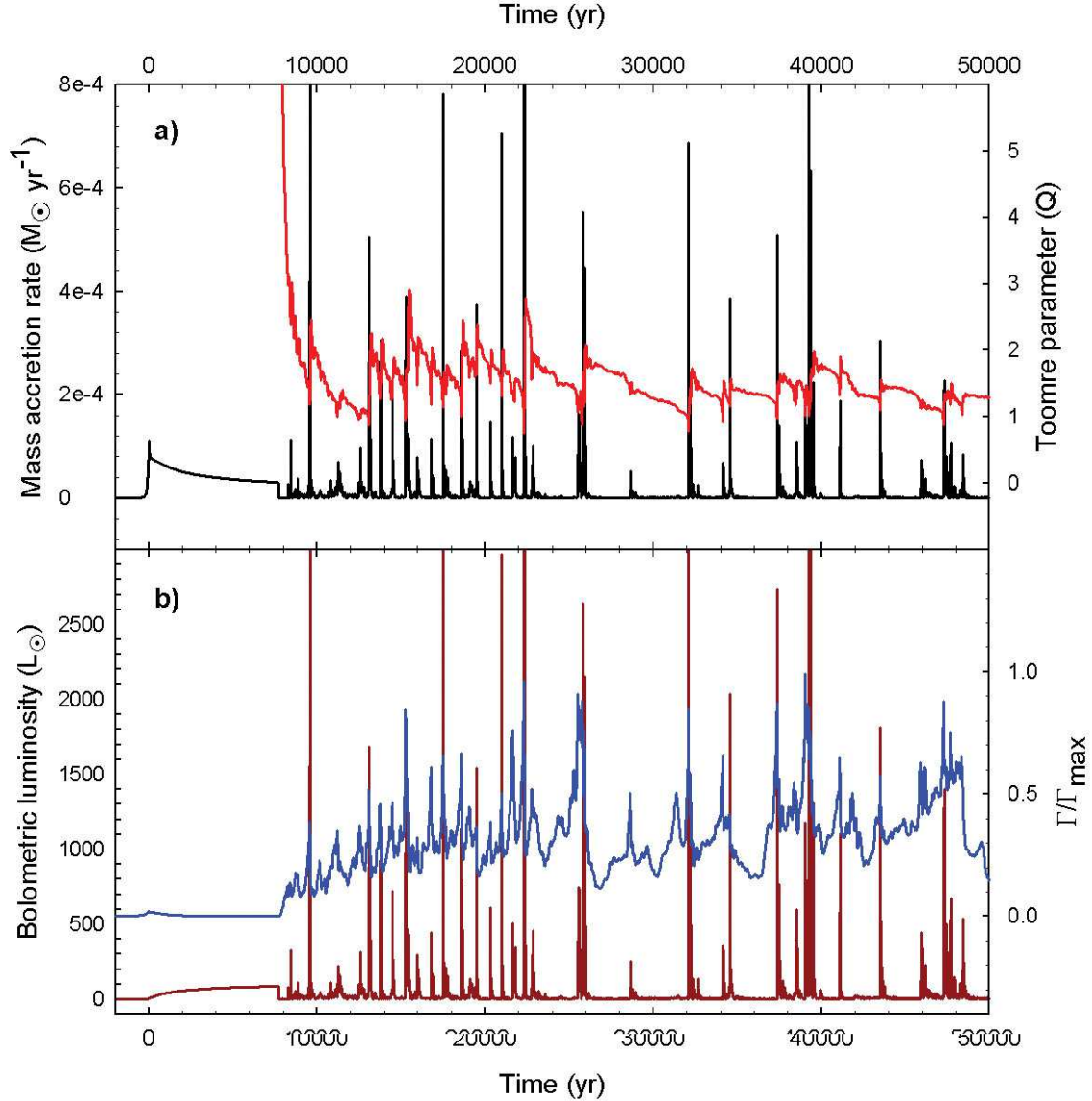}  
\caption{Periodic mass accretion outbursts.
      The temporal evolution of {\bf a)} the mass accretion rate $\dot{M}$ (the black line) 
      and the Toomre parameter $Q$ (the red line) and {\bf b)} the
      bolometric luminosity (the brown line) 
      and the normalized gravitational torque per unit mass $\Gamma/\Gamma_{\rm
      max}$ (the blue line). The horizontal axis shows the elapsed time since the formation of 
      the protostar.
      The behavior of $\dot{M}$ shows two distinct
      phases. In the early phase, $\dot{M}$ slowly declines and tends to
      approach a constant value. In the later phase, after the formation
      of the protostellar disk at $t=7700$~yr, the mass accretion occurs in periodic
      bursts.  
      The evolution of both $Q$ and $\Gamma/\Gamma_{\rm
      max}$ show a correlation with the accretion bursts.  \label{fig1}}
\end{figure*}

\begin{figure*}
 \centering
  \includegraphics[width=13 cm]{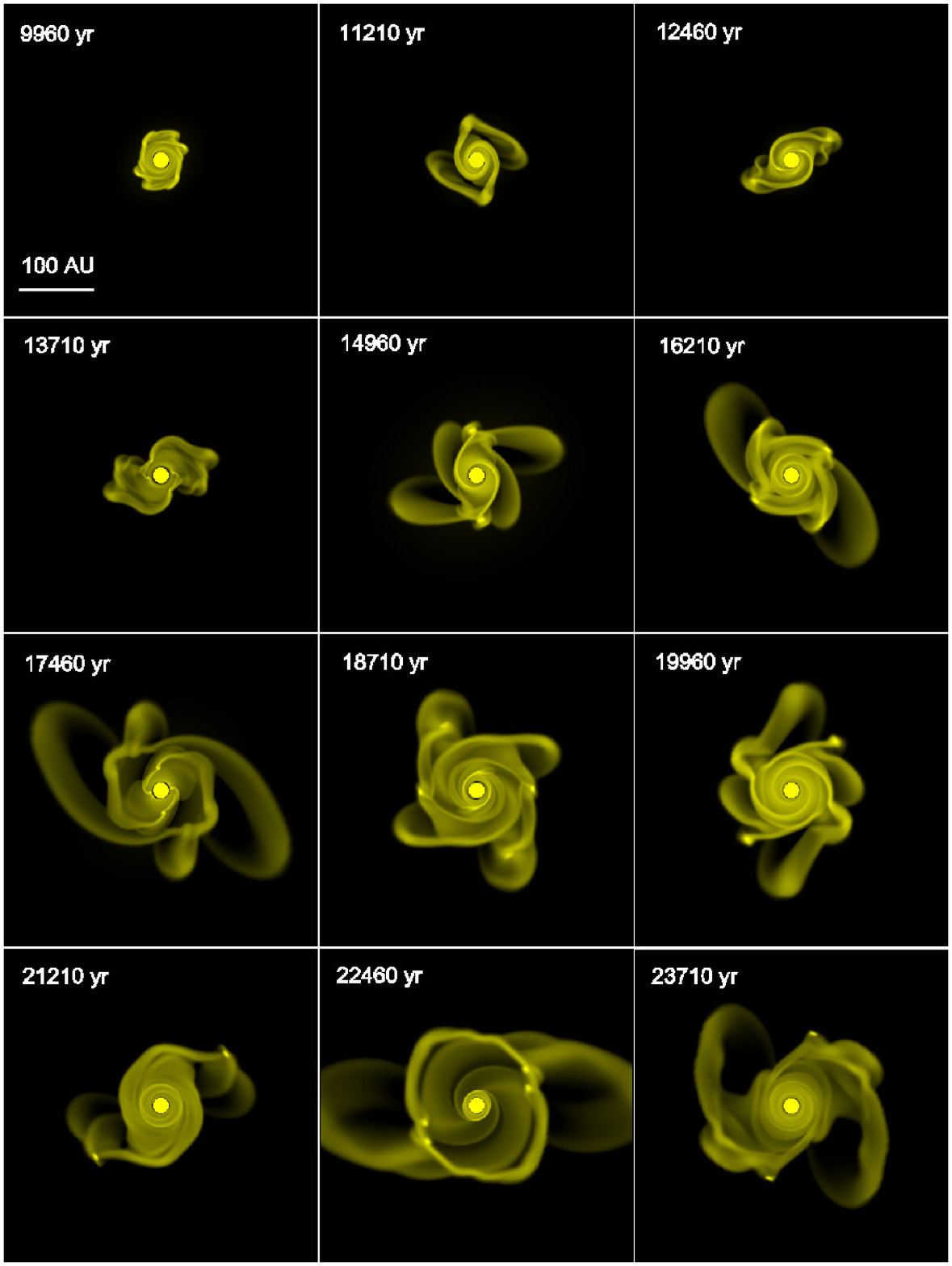}  
\caption{A sequence of gas surface density images showing the evolution
      of the protostellar disk after the formation of the protostar at $t=0$~yr. The
      disk quickly becomes unstable and develops a spiral structure with
      dominant $m=2$ and $m=4$ modes. Formation of dense protostellar/protoplanetary
      clumps within the spiral arms is evident in most images. The numbers
      in the left upper corner of each image show the elapsed time since
      the formation of the protostar.   \label{fig2}}
\end{figure*}

The cloud evolution is characterized by a slow initial gravitational
contraction and then a very rapid runaway collapse until the formation
of the central protostar.
The black line in Figure~\ref{fig1}a shows two distinct phases in the temporal evolution of
the mass accretion rate $\dot{M}$ onto the protostar.
The early behavior of $\dot{M}$ is qualitatively 
similar to that obtained in spherical collapse simulations \citep[see e.g.,][]{VB}.
Accretion shows a very rapid increase to a maximum value of 
$\dot{M}=1.0\times 10^{-4}~M_\odot$~yr$^{-1}$
at $t=0$ yr, when the
central protostar forms. Subsequently, there is a slow decline in $\dot{M}$, 
when the gas
is accreted directly onto the protostar from the inner envelope, which has
a relatively low specific angular momentum.
The second phase starts at $t=7700$~yr when
the protostellar disk forms around the protostar due to the infall
of matter from the outer envelope, which has a higher specific angular momentum.
The mass of the protostar at this stage is approximately
$M_{\rm s}=0.4~M_{\odot}$.
The matter in the disk moves on nearly
circular orbits and most of it is far from the protostar. 
Hence, the accretion rate 
onto the protostar abruptly drops down to a negligible value.
A continuous infall of matter from the envelope makes
the protostellar disk unstable to the development of spiral structure shown in Fig.~\ref{fig2}
and induces the formation of dense protostellar/protoplanetary clumps within
the arms. The even ($m=2,4$) modes are dominant in this simulation, but the odd
($m=1,3$) modes are also excited in some simulations.
Spiral arms transport angular momentum outward and mass inward \citep{Lynden}.  The
ensuing redistribution of mass
and angular momentum creates a strong centrifugal disbalance in the protostellar
disk and triggers bursts of mass accretion when dense 
protostellar/protoplanetary clumps in the inner disk are driven into
the protostar (an animation of the disk evolution can be downloaded from
{\tt http://www.astro.uwo.ca/$\sim$basu/mv.htm}). 
During this process, the mass in the protostellar disk
remains somewhat less than the mass of the protostar.
The episodes of clump infall can manifest themselves as very short 
($\le 100$~yr) but vigorous ($\dot{M}=[1-10]\times 
10^{-4}~M_{\odot}$~yr$^{-1}$) accretion bursts as is clearly seen in
Fig.~\ref{fig1}. 
During the accretion bursts, $0.01-0.05~M_{\odot}$ of gas is accreted
and the accretion luminosity may by grow many orders of magnitude.
The duration of the intervening
quiescent accretion phase with $\dot{M}=(1-10)\times 10^{-7}~M_\odot$~yr$^{-1}$
is usually $(1-3) \times 10^3$ yr. The frequency
of bursts decreases with time and the number of bursts may amount
to 15-30.  The brown line in Fig.~\ref{fig1}b shows 
a luminosity $L_{\rm bol}=G M_{\rm s} \dot{M}/R_{\rm c}$ 
(it is assumed to derive entirely from the disk accretion), where $R_{\rm c}=4
R_\odot$ is the radius of the protostar \citep{MI00}. Our $L_{\rm bol}$ 
is an upper limit to the expected observable bolometric luminosity.
It is evident that the episodes of clump infall lead to a dramatic increase
in $L_{\rm bol}$ (by a factor of up to 2000 as compared to a quiescent period) 
and the protostar may reveal itself as an FU Orionis variable.

Our calculations of the approximate Toomre parameter $Q={\tilde c}_{\rm s} \Omega/(\pi G \Sigma)$
\citep{Toomre} and the total gravitational torque per unit
mass $\Gamma$, which is the
sum of the individual torques per unit mass ($|\partial \Phi /\partial \phi|$) 
on all computational cells,
support the above scenario of episodic accretion. 
The quantity ${\tilde c}^2_{\rm s}\equiv d{\bar P}/d\Sigma$ is the squared effective
sound speed.
The $Q$ parameter may serve as an approximate stability criterion -- gas
disks are gravitationally unstable to local nonaxisymmetric perturbations 
if $Q\le 1.5-1.7$ \citep{Nelson,Boss}, while $\Gamma$ may roughly express the efficiency
of angular momentum and mass redistribution by spiral inhomogeneities
in the disk \citep{Tomley2}.  The Toomre parameter is calculated by averaging
${\tilde c}_{\rm s}$, $\Omega$, and $\Sigma$ over all computational cells. 
The red and blue lines in Fig.~\ref{fig1}a and \ref{fig1}b show the evolution of $Q$
and the normalized  gravitational torque $\Gamma/\Gamma_{\rm
max}$ after the protostellar disk formation, respectively. In the early phase of near 
constant accretion,
the matter is directly accreted onto the protostar and $Q$ is much larger than unity. 
When the protostellar disk forms, its density starts to grow due to accretion.
As a consequence, the Toomre parameter gradually  decreases below the stability limit
$Q\approx1.5$ and reaches a minimum value at the time of the accretion burst.
This strongly suggests a causal link between the gravitational instabilities and
accretion bursts.
The behavior of $\Gamma$ also shows a direct correlation with accretion
bursts. The gravitational torque gradually increases and reaches a maximum
value at the time of each accretion burst, indicating the growing efficiency
of inward mass transport before the burst.
We note that the strength of
the torque due to artificial viscosity (used in our Eulerian code to 
smooth shocks) is at least an order of magnitude smaller
than the strength of the gravitational torque associated with spiral instabilities.
Thus, the artificial viscosity cannot be responsible for the accretion
bursts. After the burst, the mass of the protostellar disk decreases
and the growth of spiral instabilities becomes temporarily suppressed,
as indicated by high values of $Q$ ($>1.5$) and a sharp decrease in $\Gamma$.
However, the continuous mass infall onto the disk from the envelope quickly
destabilizes the disk and the cycle repeats until most
of the envelope mass is accreted by the protostar.

Ambipolar diffusion, while not included in our model, is expected to 
favor the formation of clumps and subsequent burst activity
by removing magnetic support. Indeed, a model with no magnetic 
support ($\alpha = 0$) 
shows an increase in the frequency and amplitude of accretion bursts (Vorobyov
\& Basu, in preparation).
Further magnetic effects such as magnetic braking \citep{Krasnopolsky}
and magnetorotational instability \citep{Fromang} may become important 
in the late accretion phase, 
but can only be studied in a future three-dimensional model.

\section{Conclusions}
\label{concl}
Rotating protostellar cores show two distinct phases in the temporal evolution of the mass accretion rate
$\dot{M}$ onto the protostar.  The early behavior of $\dot{M}$ is qualitatively similar to that obtained in 
spherical collapse simulations. Accretion shows a very rapid increase to a maximum, 
when the central protostar forms, and a subsequent slow decline, when the gas is accreted directly onto the protostar 
from the inner envelope. The second phase starts when the protostellar disk forms around the protostar 
due to the infall of matter from the outer envelope with higher specific angular momentum. 
In this phase, $\dot{M}$ is characterized by very short ($<100$~yr) but vigorous
($\dot{M}=[1-10]\times 10^{-4}~M_{\odot}$~yr$^{-1}$) accretion bursts, 
which are intervened with longer periods ($\sim 10^3$~yr) of quiescent accretion. 

We have demonstrated that
the repeating accretion bursts reflect a basic 
self-regulation mechanism that is inherent to self-gravitating 
rotating protostellar disks. 
We emphasize that it is the ongoing infall of matter from the 
protostellar envelope that continually destabilizes the disk
and causes it to periodically dump significant amounts of matter
onto the protostar while transferring excess angular momentum to the 
envelope.
The effect of additional support due to a frozen-in 
supercritical magnetic field is to moderate the burst activity 
but not suppress it.
Recent high resolution Hubble Space Telescope observations 
of the central regions ($\le$ a few hundred pc) of Seyfert galaxies \citep{Regan}
also reveal rich spiral structures.  We suggest
that a self-regulation mechanism similar to that in our model
may operate on galactic scales
and be responsible for periodic nuclear activity in at least some
Seyfert galaxies.

\begin{acknowledgements}
This research was supported by the Natural Sciences and Engineering
Research Council of Canada. EIV gratefully acknowledges support
from a CITA National Fellowship. 
\end{acknowledgements}


\begin{thebibliography}{}


\bibitem[Basu(1997)]{Basu2} 
Basu, S. 1997, ApJ, 485, 240  

\bibitem[Basu \& Ciolek(2004)]{Basu3}
Basu, S., \& Ciolek, G. E. 2004, ApJ, 607, L39

\bibitem[Bate et al.(2003)]{Bate2}
Bate, M. R., Bonnell, I. A., \& Bromm, V. 2003, MNRAS, 339, 577

\bibitem[Bell \& Lin(1994)]{Bell}
Bell, K. R., \& Lin, D. N. C. 1994, ApJ, 427, 987 

\bibitem[Binney \& Tremaine(1987)]{BT}
Binney, J., \& Tremaine, S. 1987, Galactic~Dynamics,
(Princeton Univ. Press, Princeton), 96

\bibitem[Bonnell \& Bastien(1992)]{Bonnell}
Bonnell, I., \& Bastien, P. 1992, ApJ, 401, L31 

\bibitem[Boss(1998)]{Boss} 
Boss, A. P. 1998, ApJ, 503, 923

\bibitem[Boss(2003)]{Boss2}
Boss, A. P. 2003, ApJ, 599, 577  

\bibitem[Clarke et al.(1990)]{Clarke}
Clarke, C. J., Lin, D. N. C., \& Pringle, J. E. 1990, MNRAS, 242, 439 

\bibitem[Corder et al.(2004)]{Sargent}
Corder, S., Eisner, J., \& Sargent, A. 2005, ApJ, 622, 133

\bibitem[Fromang et al.(2005)]{Fromang} 
Fromang, S., Balbus, S. A., Terquem, C., \& De Villiers, J.-P. 2005, 
\apj, 616, 364

\bibitem[Fukagawa et al.(2004)]{Fukagawa}
Fukagawa, M., Hayashi, M., Tamura, M. et al. 2004, ApJ, 605, L53

\bibitem[Kenyon et al.(1990)]{Kenyon2}
Kenyon, S. J.,  Hartmann, L. W., Strom, K. M., \& Strom, S. E. 1990,  AJ,
99, 869 

\bibitem[Kenyon \& Hartmann(1995)]{Kenyon}
Kenyon, S. J., \& Hartmann, L. W. 1995, ApJS, 101, 117 

\bibitem[Krasnopolsky \& K{\"o}nigl(2002)]{Krasnopolsky}
Krasnopolsky, R., \& K{\"o}nigl, A. 2002, ApJ, 580, 987

\bibitem[Larson(2003)]{Larson}
Larson, R. B. 2003, Rep. Prog. Phys., 66, 1651

\bibitem[Laughlin \& Bodenheimer(1994)]{LB}
Laughlin, G., \& Bodenheimer, P. 1994, ApJ, 436, 335 

\bibitem[Lin \& Papaloizou(1985)]{Lin}
Lin, D. N. C. \& Papaloizou, J. C. B. 1985, in~Protostars~and~Planets~II,
ed. D. C. Black, M. C. Matthews, Eds. (Univ. Ariz. Press, Tucson, 1985), 981 

\bibitem[Lynden-Bell \& Kalnajs(1972)]{Lynden}
Lynden-Bell, D., \& Kalnajs, A. J. 1972, MNRAS, 157, 1 

\bibitem[Masunaga \& Inutsuka(2000)]{MI00}
Masunaga, H., \& Inutsuka, S. 2000, ApJ, 531, 350

\bibitem[Mejia et al.(2005)]{Mejia}
Mejia, A. C., Durisen, R. H., Pickett, M. K., \& Cai, K. 2005, ApJ, 619,
1098

\bibitem[Nakamura \& Hanawa(1997)]{Nakamura}
Nakamura, F., \&  Hanawa, T. 1997, ApJ, 480, 701 

\bibitem[Nelson et al.(1998)]{Nelson}
Nelson, A. F., Benz, W., Adams, F. C., \& Arnett, D. 1998, ApJ, 502, 342

\bibitem[Regan \& Mulchaey(1999)]{Regan}
Regan, M. W., \& Mulchaey, J. S. 1999, AJ, 117, 2676 

\bibitem[Shakura \& Sunyaev(1973)]{Shakura}
Shakura, N. I., \& Sunyaev, R. A. 1973, A\&A, 24, 337 

\bibitem[Shu \& Li(1997)]{Shu}
Shu, F. H., \& Li, Z.-Y. 1997, \apj, 475, 251

\bibitem[Stone \& Norman(1992)]{SN}
Stone, J. M., \& Norman, M. L. 1992, ApJS, 80, 753 

\bibitem[Tomley et al.(1994)]{Tomley2}
Tomley, L., Steiman-Cameron, T. Y., \& Cassen, P. 1994, ApJ, 422, 850 

\bibitem[Toomre(1981)]{Toomre}
Toomre, A. 1981, in~The~Structure~and~Evolution~of~Normal~Galaxies, ed.
S. M. Fall, D. Lynden-Bell, (Cambridge Univ. Press, Cambridge), 111

\bibitem[Truelove et al.(1998)]{Truelove}
Truelove, J. K., Klein, R. I., McKee, C. F., Holliman II, J. H., Howell,
L. H., Greenough, J. A., \& Woods, D. T. 1998, ApJ, 495, 821 

\bibitem[Vorobyov \& Basu(2005)]{VB}
Vorobyov, E. I., \& Basu, S. 2005, MNRAS, 360, 675 


\end{thebibliography}
\end{document}